# Controlling the electronic structure of graphene using surface-adsorbate interactions

*Piotr Matyba †, Adra V. Carr †, Cong Chen †, David L. Miller ‡, Guowen Peng ||, Stefan Mathias §, Manos Mavrikakis ||, Daniel S. Dessau ⊥, Mark W. Keller ‡, Henry C. Kapteyn, and Margaret Murnane †.*

† Department of Physics and JILA, University of Colorado and NIST, Boulder, Colorado 80309, United States

‡ National Institute of Standards and Technology (NIST), 325 Broadway, Boulder, Colorado 80305, United States

§ Department of Physics and Research Center OPTIMAS, University of Kaiserslautern, 67663 Kaiserslautern, Germany

⊥ Department of Physics, University of Colorado, Boulder, Colorado 80309-0390, USA

|| Department of Chemical and Biological Engineering, University of Wisconsin-Madison, Madison 53706, United States

Abstract

We show that strong coupling between graphene and the substrate is mitigated when 0.8 monolayer of Na is adsorbed and consolidated on top graphene-on-Ni(111). Specifically, the π state is partially restored near the K-point and the energy gap between the π and π* states reduced to 1.3 eV after adsorption, as measured by angle-resolved photoemission spectroscopy. We show that this change is not caused by intercalation of Na to underneath graphene but it is caused by an electronic coupling between Na on top and graphene. We show further that graphene can be decoupled to a much higher extent when Na is intercalated to underneath graphene. After intercalation, the energy gap between the π and π* states is reduced to 0 eV and these states are identical as in freestanding and n-doped graphene. We conclude thus that two mechanisms of decoupling exist: a strong decoupling through intercalation, which is the same as one found using noble metals, and a weak decoupling caused by electronic interaction with the adsorbate on top.

Graphene is a one atom thick sheet of carbon with unique electronic properties that enable novel applications in optoelectronic devices [1]. The linear dispersion of the π and π* states near the corner of the graphene Brillouin zone, where these states intersect the Fermi level and form the Dirac cone [1, 2], facilitates the zero effective mass and extremely high mobility of charge careers, ballistic charge transport, Berry's phase, and the anomalous quantum Hall effect [3-5]. These electronic properties, which are desired for applications, can be obliterated when coupling of graphene to the substrate is strong, or when adsorbates/contaminants reside on the surface [6-8]. Specifically, when graphene is grown on Ni(111) [Gr/Ni(111)] the coupling to the substrate is very strong. The π state is hybridized with the substrate states causing that the Dirac cone is destroyed [8-11]. Although these hybridized states split into a gamut of states of which some may intersect at the K-point, forming Dirac cones at binding energies below the Fermi level [12, 13], the properties of these states are not well understood and difficult to interrogate. Restoring the Dirac cone at the K-point, which potentially can open the way to applications, can achieved by intercalating noble metal atoms to underneath Gr/Ni(111) [8-11]. These atoms form an atomic decoupler that liberates the layer of graphene from the substrate. Although similar effect should be achieved using alkali atoms, experiments to date suggested that Na and K cause only an incomplete decoupling, reducing the π-to-π* energy gap to ≈1.3 eV and not restoring the linear character of these states at the K-point [14, 15].

In this paper, we show that this incomplete decoupling results from the electronic interaction of Gr/Ni(111) with alkali atoms adsorbed on top rather than from intercalation. We show that the intercalation causes a stronger decoupling, restoring the π and π* states near the K-point, and reducing the π-to-π* energy gap to ≈0 eV. We use angle-resolved photoemission spectroscopy (ARPES) and density functional theory (DFT) to show dynamic changes in the energy and dispersion of states near the K-point induced by adsorption and intercalation of Na. These changes reveal very subtle aspects of coupling between graphene and the substrate and show that decoupling through intercalation of Na is indeed identical as one through intercalation of noble metals [8-11]. Graphene is detached from the substrate and hybridization with the substrate is not possible after intercalation of Na. The new ability to control the band gap in graphene using electronic interactions, rather than intercalation, suggests that it should be possible to manipulate the electronic structure on fast time-scales, since the extent of electronic coupling can be controlled using various photoexcitation schemes [16, 17].

The high quality epitaxial Gr/Ni(111) used in this study was produced by dissociating $C_2H_4$ gas on Ni(111) single crystal films at 900 K and under $C_2H_4$ pressure of $10^{-6}$ Torr. Na was adsorbed under a pressure of ≈$10^{-10}$ Torr and at room temperature using a commercial Na source. Intercalation of Na was achieved by annealing the sample at temperatures in the range of 300 to 450 K followed by rapid cooling to room temperature. ARPES was performed using VUV radiation at 40.8 eV from a He discharge lamp (Specs UVS300) and a hemispherical analyzer (Specs Phoibos 100) [18]. The morphology of Gr/Ni(111) was examined using ARPES, low-energy electron diffraction (LEED), and kinetics of Na intercalation and Na oxidation under exposure to $O_2$. DFT calculations were performed using the VASP code [19] based on spin-polarized DFT. The projector augmented wave potentials were used to model electron-ion interactions [20, 21] and Van der Waals dispersion forces were accounted for using the vdW-DF approach [19, 22-24]; details are in supplementary information.

Due to the size of atomic Na [25], the intercalation into graphene is facilitated by defects and grain boundaries, requiring high mobility of Na and consequently elevated temperatures. Once intercalated, we found that Na can escape through these defects at high temperatures, although energetics (as discussed later) indicates that Na should remain under Gr/Ni(111). This effect is due to the high mobility of Na ions at elevated temperatures and a low energy barrier [13], which depends strongly on the morphology of graphene. In experiments, we detected Na on both sides of graphene immediately after adsorption and further temperature treatments could not increase the fraction of intercalated Na. Defects produced during formation of graphene by dissociation of $C_2H_4$ on Ni(111) cause that graphene is permeable to Na, i.e. not able to confine Na either on top or underneath. Further ARPES and LEED studies confirmed that graphene of poor morphology was permeable to Na, causing that highly mobile sodium can easily penetrate from top to underneath and reverse. The oxidation of adsorbed Na under exposure to $O_2$ in UHV was used as an indicator of intercalation since oxidation can take place only when Na remains on top of graphene, owing to the size of $O_2$ molecule [26].

The improved morphology was achieved by dissolving graphene into the substrate at high temperatures (>1000 K) followed by slow (≈2 hrs) precipitation of C onto the surface, forming a new graphene. The solubility of C in Ni, which is ≈3 times higher at 1100 K than at room temperature [27], makes it possible to control the formation of graphene at the surface of Ni(111) by means of cooling rates. Graphene formed at lower rates, through precipitation of carbon rather than dissociation of $C_2H_4$, showed much better morphology. This precipitation procedure was vital for achieving large domains of graphene and the ability to control the adsorbed (or intercalated) Na. We observed that the intercalation of Na into the highest quality Gr/Ni(111) required elevated temperatures close to the Na desorption threshold.

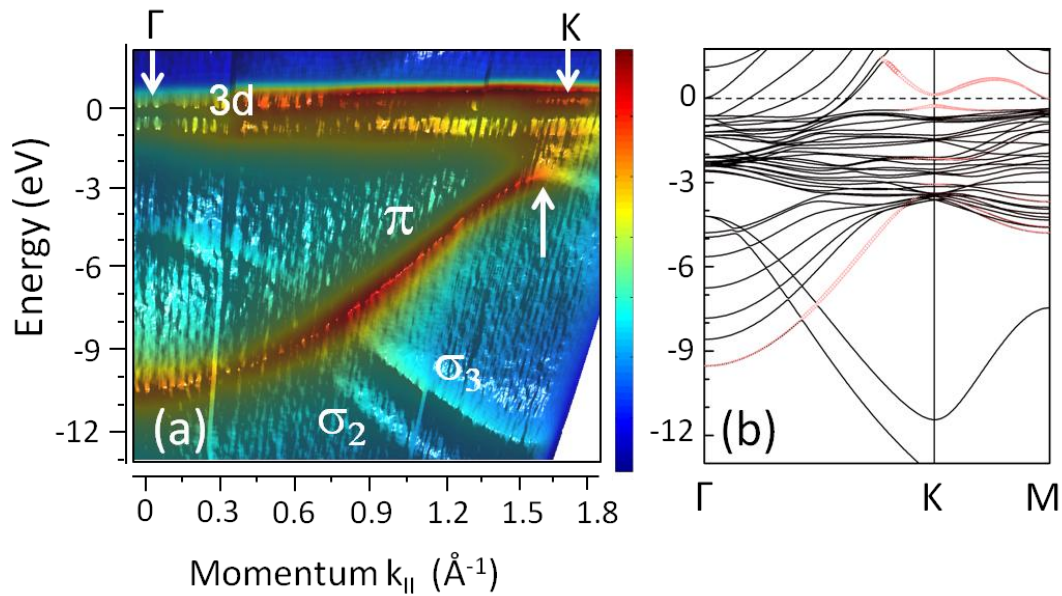

**Fig. 1.** (a) Electronic structure of pristine Gr/Ni(111) system along the ΓK direction in the Brillouin zone; π, $\sigma_2$ and $\sigma_3$ states of graphene and the nickel 3d band are indicated with symbols, high symmetry points of the Brillouin zone (Γ, K), and the π state 'cut-off' are indicated with vertical white arrows. (b) Calculated majority band structure of Gr/Ni(111) (details in the text), graphene $2p_z$ contributions are highlighted in red.

The dispersion of the π state near the K-point was previously probed along the ΓK [11, 12, 14] and p-ΓK [12, 13, 28] directions and emerged as a sensitive probe of the electronic environment of graphene. In Gr/Ni(111), we measured the state dispersion along the ΓK direction, as shown in Fig. 1(a), in the energy range from the Fermi level down to -13 eV. In agreement with previous experiments, we find that the π state intersects the Ni(111) d band near the K-point and the state dispersion deviates from linear near the intersection. The state is downshifted by 2 eV at the Γ-point and downshifted by 2.8 eV near the K-point, as compared to freestanding graphene [11, 14, 29]. The state extends all the way to the K-point but its intensity is strongly diminished at momenta ≥1.5 Å$^{-1}$ (see supplementary material), in agreement with previous measurements [12, 13, 28].

The experimentally measured electronic structure in Fig. 1(a) is dominated by the strongest features, which are predominantly the d band, π, $\sigma_2$, and $\sigma_3$ states. Therefore, further insight into the electronic structure of Gr/Ni(111) is given by DFT calculations in Fig. 1(b). The electronic structure calculated for the most stable top-fcc configuration and for the majority of spin gives access to lower intensity states that are not accessible in the measurements. Since the calculations omit the spectral weights or matrix elements relevant for ARPES, they show the energy and dispersion of all states, not showing the relative intensities. This is vital for simultaneous identification of weak and strong features near the Fermi level.

The calculations show that the π state and the d band are hybridized near the K-point [14]. Close to the Fermi level, where the density of Ni(111) states is lower, the π state is hybridized to a lower degree and can be identified in Fig. 1(b) (indicated with red line). At lower energies below the Fermi level down to -2.8 eV the hybridization causes that the π state and the d band are split into a dense manifold of states and modified by hybridization effects, as visible in Figs. 1(a) (an enlarged spectrum is shown in supplementary material, Fig. S1). The intensity of these hybridized states is low along the ΓK direction, presumably due to matrix element effects induced by hybridization [30-32], but it is possible to probe these states using photons of higher energy and/or other cuts through the Brillouin zone [12, 13]. High resolution ARPES along p-ΓK direction showed that these states may intersect, forming Dirac cones near the K-point at binding energies below the Fermi level [12, 13].

We observed that the π* state is modified by hybridization effects and shifted in energy to above the Fermi level in Fig. 1(b). In a close-up shown in Fig. 2(a) this state remains unpopulated and is undetectable in ARPES. The adsorption and consolidation of 0.8 monolayer of Na on top causes that surface is n-doped and the Fermi level shifts upwards such that the π* state becomes populated, in Fig. 2(b). These visible changes of the π* state are clearly due to n-doping but the visible changes of the π state in the same spectrum cannot be explained as due to n-doping. The dispersion of the π state changes towards linear in the entire range of the momenta shown in Fig. 2(b). The state maximum, indicated roughly by yellow arrows (↑) in figure, shows up at a higher energy and surpasses the hybridized manifold visible in Fig. 2(a); the π-to-π* energy gap after adsorption is lowered to ≈1.3 eV.

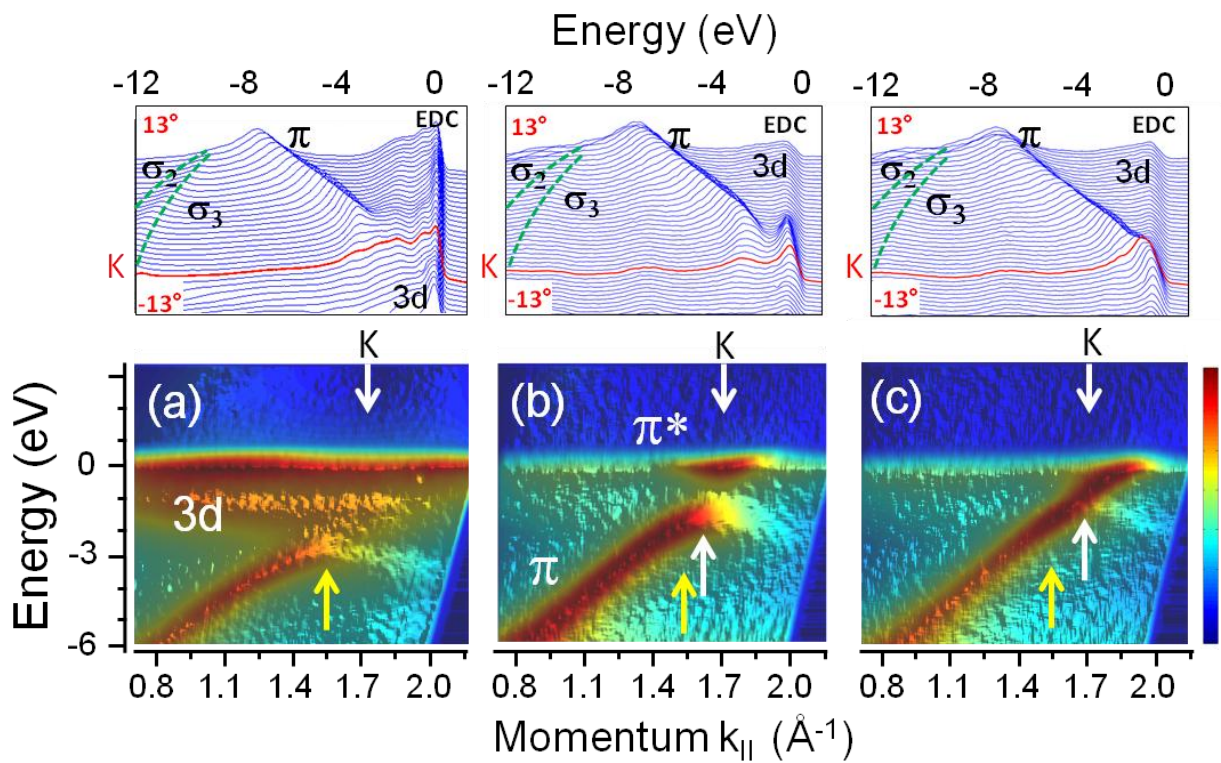

**Fig. 2.** (a) Band structure of graphene on Ni(111), K-point is indicated with vertical white arrows (↓). The insets above show the energy distribution curves (EDCs) near the K-point (K-point EDS is indicated as red) as a function of detection angle. The EDCs energy range is from -12 eV to 1 eV. (b) Same as (a) but after adsorption and consolidation of 0.8 monolayer Na on top; the minimum of the π* state is visible at the Fermi level. Yellow and white vertical arrows (↑) indicate roughly the maximum of the π state, as measured in the pristine and Na adsorbed samples, respectively. (c) Same as (b) after further annealing to intercalate Na to underneath graphene.

Previous studies that showed that the π-to-π* energy gap in Gr/Ni(111) is ≈1.3 eV after adsorption of Na (or K) [14, 15] overlooked changes in the π state momentum and intensity. In their interpretation, the π-to-π* energy gap was reduced as a result of intercalation of the adsorbates to underneath the graphene layer and the concomitant decoupling, although no proof of intercalation was shown. Only changes of the π state energy, momentum, and intensity near the K-point can confirm intercalation and decoupling [33]. The intercalation cannot be confirmed by angle-resolved XPS because the angular dependence of the XPS lines from Na (K) covered Gr/Ni(111) is caused by the formation of thick Na (K) islands on top [26, 34-36] rather than intercalation.

Our experimental results supported by DFT calculations (discussed later) exclude intercalation as the origin of changes shown in Fig. 2(b). We also exclude intercalation at isolated locations causing patches of decoupled and still-coupled graphene. Such partial intercalation would cause emergence of two π states in ARPES spectra, one from intercalated and one from non-intercalated areas as observed in reference [13]. We show below that intercalation of Na causes complete decoupling and closing the π-to-π* energy gap by breaking down the hybridization near and at the K-point.

It has been shown several times to date that large atoms break down the hybridization when intercalated underneath Gr/Ni(111). However, the mechanism of intercalation and decoupling by alkali metals is still under debates [9, 10, 13, 26, 28, 34, 35, 37-41]. Intercalation of Na into graphene on 4H-SiC(0001) and 6H-SiC(0001) at elevated temperatures is possible only at grain boundaries and structural defects [35, 41, 42]. Accordingly, we show in Fig. 2(c) ARPES spectrum recorded after careful annealing of the sample at and above 400 K for 2-3 minutes to enable intercalation. The sample was kept in front of the detector at the same position as in Fig. 2(b) at all times (proof of intercalation is shown in supplementary material).

Notably, the π state is restored at the K-point, the π-to-π* energy gap is reduced to ≈0 eV, the Dirac cone is intact along the ΓK direction, and the Fermi velocity, which is a good indicator of the substrate coupling [43], is increased towards the value of freestanding graphene (≈1.6×10$^6$ m/s). The data clearly shows that graphene is decoupled from the substrate and that the π state is not hybridized with the d band. The mechanism of decoupling when an adsorbate is intercalated is quite trivial, as discussed elsewhere [9, 10, 13, 26, 28, 34, 35, 37-41]. The adsorbate forms an atomic decoupler that physically separates graphene and the substrate, not allowing neither for hybridization nor electronic coupling. The situation is far more complex when Na remains on top. Figures 3(b), 3(c), and 3(d) show calculated electronic structures of Gr/Ni(111) without Na, with Na on top, and with Na intercalated, respectively.

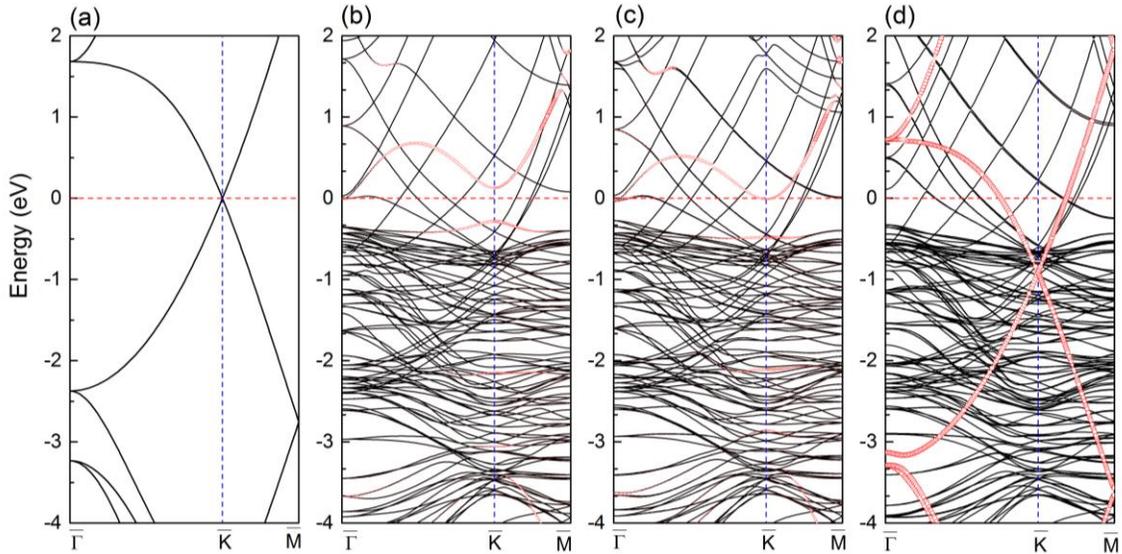

**Fig. 3.** Band structures for (a) freestanding graphene, (b) Gr/Ni(111), (c) Na/Gr/Ni(111), and (d) Gr/Na/Ni(111) in a (2×2) supercell, corresponding to a Na coverage of 0.75 ML. The contributions of graphene 2p$_z$ states are highlighted in red in panels (b)-(d).

Since the surface work function is lowered after adsorption, the energies of all states in Fig. 3(c) should be downshifted compared to Fig. 3(b). Indeed, the bottom of the π* state [indicated as red line above the Fermi level in Figs. 3(b) and 3(c)] is shifted to below the Fermi level, as confirmed experimentally in Fig. 2(b). The hybridized part of π state, near the K-point [red line below the Fermi level in Figs. 3(b) and 3(c)], is also downshifted. However, the other hybridized states below the Fermi level remain virtually unchanged, suggesting that some screening effects take place and keep them at a fixed energy throughout the adsorption. The intensity (i.e. spectral weight) of these hybridized states is strongly diminished after adsorption [44], suggesting that this screening is responsible for the weak decoupling which we see in Fig. 2(b).

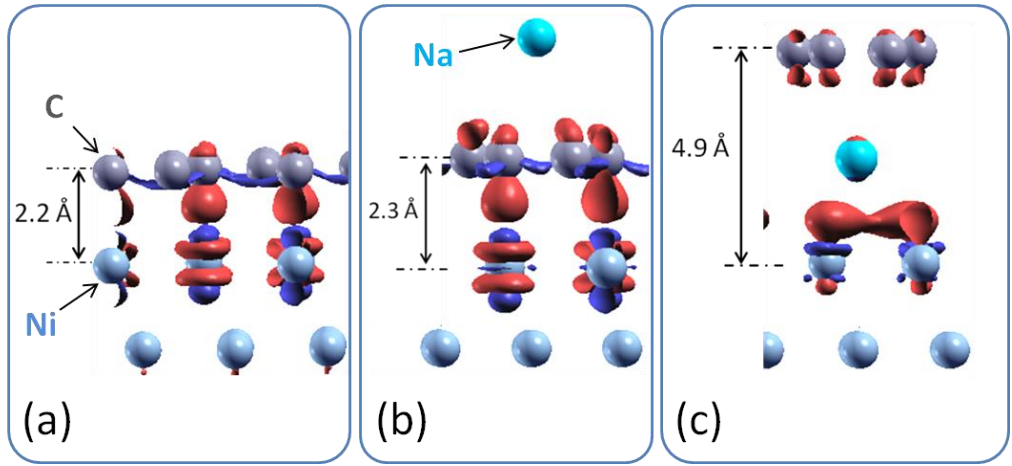

**Fig. 4.** Charge density difference plots, using an isosurface of ±0.02 e/Å$^3$ for (a) Gr/Ni(111), (b) Na/Gr/Ni(111), and (c) Gr/Na/Ni(111). Charge density accumulation is shown in red and depletion in blue. The adsorption energy of Na on Gr/Ni is -1.00 eV at a Na coverage of 1 ML, referenced to the total energies of atomic Na and Gr/Ni(111).

The plots of charge density difference before and after Na adsorption, shown in Figs. 4(a) and 4(b), give more physical intuition about the decoupling. The d band and the π state intersect near the K-point and hybridize [cf. Fig. 4(b)], yielding a charge density redistribution that minimize the total energy of Gr/Ni(111). In the top-fcc (energetically most stable) arrangement [in Fig. 4(a)] Bader charge analysis shows that a net charge of 0.10 e (e is the elementary charge) is donated from the substrate to the graphene 2p$_z$ orbitals, which give rise to the hybridized π state. With Na on top, the energetics are different. Na donates a net charge of 0.36 e per atom to graphene (to the π and π* states) and the substrate is well screened from the adsorbate receiving no charge. These donated electrons populate the orbitals involved in hybridization, lowering their propensity of bonding to Ni(111). As a result, graphene is 2.30 Å above the substrate (lifted by 0.1 Å) after Na adsorption and the charge transfer from the substrate to graphene drops to only 0.05 e per atom. Detailed energetics of decoupling are complex, but they can be summarized as follows: The hybridization is energetically less favorable when the adsorbate on top provides a rich source of electrons. The adsorbate causes thus weak decoupling by means of charge rearrangements and electronic screening, which is visible in the experiments in Fig. 2(b). In agreement with this scenario, we anticipate that a higher coverage (a higher net charge transfer) or higher electronegativity of the adsorbate would yield even stronger decoupling, which should be consistent with previous measurements [15].

We show the charge density difference in intercalated Gr/Ni(111) [in Fig. 4(c)] and the corresponding band structure [in Fig. 3(c)]. DFT calculations show that Na on top of graphene is energetically less stable than intercalated, e.g. by 1.26 eV at a Na coverage of 0.75 monolayer. Therefore, Na tends to intercalate through grain boundaries, as predicted in ref. [13], but details of the energy barrier for intercalation are complex and depend on the size of graphene and surface morphology. In the intercalated case graphene is lifted up 4.94 Å above the substrate and is completely decoupled. The calculated band structure and the experiments in Fig. 2(c), show clearly that the hybridized states regained the pristine character. They are akin now to states of freestanding and n-doped graphene; cf. Fig. 3(a).

With this result, we turn our attention to studies of adsorbates on Gr/Ni(111) available in the literature. Past work claimed that intercalation of Na (and K) can reduce the π-to-π*energy gap to ≈1.3 eV and overlooked changes in the dispersion of the π state near the K-point [14, 15]. Here, we obtain identical results with no intercalation and we show that intercalation produces quasi-freestanding graphene, consistent with studies using other intercalants [9, 10, 13, 26, 28, 34, 35, 37-41], but not Na. To understand why the past works did not achieve the same result as in Fig. 2(c), we performed intercalation studies using Gr/Ni(111) of a lower quality. We were able to achieve relatively good control over the concentration of defects by introducing oxide impurities into the Ni(111) substrate and by performing the growth at lower temperatures (and with no reforming steps) producing impurities of amorphous carbon [45]. We found that consistent and repeatable decoupling through intercalation of Na was virtually impossible whenever LEED and ARPES measurements indicated low quality graphene, with a large number of defects and presumably smaller domains. In some cases, we found that mobility of the adsorbate at lower temperatures was sufficient to enable the intercalation. However, results of the band structure measurements were inconsistent from experiment to experiment, often indicating that Na populates in fact both sides of the graphene sheet. Graphene comprising large domain is a prerequisite for the successful intercalation and trapping Na atoms underneath.

In summary, we show that the presence of Na atoms on top of Gr/Ni(111) reduces hybridization by means of charge transfer and screening, as evident from restoration of the π state near the K-point and a decrease in the π-to-π* energy gap to 1.3 eV. Subsequent intercalation reinstates the π state at the K-point (removes hybridization) and restores the Dirac cone. We anticipate that the mechanism of controlling the extent of coupling by an adsorbate through a charge transfer rather than intercalation opens possibility of using laser excitations of adsorbates to control this coupling. This technique might allow for ultrafast switching in the spirit of previous work on noble metal surfaces [46, 47]. Potentially, a similar scheme can be applied to other graphene systems making it useful for novel optoelectronic devices.

The authors gratefully acknowledge support from the National Science Foundation Physics Frontiers Center Program. DD acknowledges support from grant DOE-BES DE-FG02-03ER46066. PM acknowledges fellowship from the Swedish Research Council (Vetenskapsrådet). Work at UW was partially supported by *Department of Energy-Basic Energy Sciences* (DOE-BES) and by the Air Force Office of Scientific Research under Basic Research Initiative grant AFOSR FA9550-12-1-0481. The computational work was performed in part using supercomputing resources from the following institutions: EMSL, a National scientific user facility at Pacific Northwest National Laboratory (PNNL); the Center for Nanoscale Materials (CNM) at Argonne National Laboratory (ANL); and the National Energy Research Scientific Computing Center (NERSC). EMSL is sponsored by the Department of Energy's Office of Biological and Environmental Research located at PNNL. CNM and NERSC are supported by the U.S. Department of Energy, Office of Science, under Contracts DE-AC02-06CH11357 and DE-AC02-05CH11231, respectively.

References

1. Geim, A.K. and K.S. Novoselov, *The rise of graphene.* Nature Materials, 2007. **6**(3): p. 183-191.

2. Zou, K., X. Hong, and J. Zhu, *Effective mass of electrons and holes in bilayer graphene: Electron-hole asymmetry and electron-electron interaction.* Physical Review B, 2011. **84**(8).

3. Zhang, Y.B., et al., *Experimental observation of the quantum Hall effect and Berry's phase in graphene.* Nature, 2005. **438**(7065): p. 201-204.

4. Novoselov, K.S., et al., *Two-dimensional gas of massless Dirac fermions in graphene.* Nature, 2005. **438**(7065): p. 197-200.

5. Novoselov, K.S., et al., *Unconventional quantum Hall effect and Berry's phase of 2 pi in bilayer graphene.* Nature Physics, 2006. **2**(3): p. 177-180.

6. Bostwick, A., et al., *Observation of Plasmarons in Quasi-Freestanding Doped Graphene.* Science, 2010. **328**(5981): p. 999-1002.

7. Ohta, T., et al., *Controlling the electronic structure of bilayer graphene.* Science, 2006. **313**(5789): p. 951-954.

8. Walter, A.L., et al., *Electronic structure of graphene on single-crystal copper substrates.* Physical Review B, 2011. **84**(19): p. 195443-195443.

9. Varykhalov, A., et al., *Electronic and Magnetic Properties of Quasifreestanding Graphene on Ni.* Physical Review Letters, 2008. **101**(15): p. 066804-066808.

10. Varykhalov, A., et al., *Effect of noble-metal contacts on doping and band gap of graphene.* Physical Review B, 2010. **82**(12): p. 121101-121105.

11. Haberer, D., et al., *Tunable Band Gap in Hydrogenated Quasi-Free-Standing Graphene.* Nano Letters, 2010. **10**(9): p. 3360-3366.

12. Varykhalov, A., et al., *Intact Dirac Cones at Broken Sublattice Symmetry: Photoemission Study of Graphene on Ni and Co.* Physical Review X, 2012. **2**(4): p. 041017-041027.

13. Park, Y.S., et al., *Quasi-Free-Standing Graphene Monolayer on a Ni Crystal through Spontaneous Na Intercalation.* Physical Review X, 2014. **4**(4): p. 031016-031025.

14. Gruneis, A. and D.V. Vyalikh, *Tunable hybridization between electronic states of graphene and a metal surface.* Physical Review B, 2008. **77**(19): p. 193401-193401.

15. Nagashima, A., N. Tejima, and C. Oshima, *Electronic States of the Pristine and Alkali-Metal-Intercalated Monolayer Graphite/Ni(111) Systems.* Physical Review B, 1994. **50**(23): p. 17487-17495.

16. Petek, H., et al., *Real-time observation of adsorbate atom motion above a metal surface.* Science, 2000. **288**(5470): p. 1402-1404.

17. Petek, H., et al., *Electronic relaxation of alkali metal atoms on the Cu(111) surface.* Surface Science, 2000. **451**(1-3): p. 22-30.

18. Equipment is identified in this paper only in order to adequately specify the experimental procedure. Such identification does not imply endorsement by the National Institute of Standards and Technology (NIST), nor does it imply that the equipment identified is the best available for the purpose.

19. Kresse, G. and J. Furthmüller, *Efficient iterative schemes for ab initio total-energy calculations using a plane-wave basis set.* Phys. Rev. B, 1996. **54**: p. 11169-11186.

20. Blochl, P.E., *PROJECTOR AUGMENTED-WAVE METHOD.* Physical Review B, 1994. **50**(24): p. 17953-17979.

21. Kresse, G. and D. Joubert, *From ultrasoft pseudopotentials to the projector augmented-wave method.* Physical Review B, 1999. **59**(3): p. 1758-1775.

22. Dion, M., et al., *Van der Waals density functional for general geometries.* Physical Review Letters, 2004. **92**(24): p. 246401-246405.

23. Klimes, J., D.R. Bowler, and A. Michaelides, *Chemical accuracy for the van der Waals density functional.* Journal of Physics-Condensed Matter, 2010. **22**(2): p. 022201-022206.

24. Klimes, J., D.R. Bowler, and A. Michaelides, *Van der Waals density functionals applied to solids.* Physical Review B, 2011. **83**: p. 195131-195144.

25. Petrovic, M., et al., *The mechanism of caesium intercalation of graphene.* Nature Communications, 2013. **4**.

26. Dedkov, Y.S., et al., *Graphene-protected iron layer on Ni(111).* Applied Physics Letters, 2008. **93**(2): p. 022509-022513.

27. Dunn, W.W., R.B. McLellan, and W.A. Oates, *SOLUBILITY OF CARBON IN COBALT AND NICKEL.* Transactions of the Metallurgical Society of America, 1968. **242**(10): p. 2129.

28. Papagno, M., et al., *Large Band Gap Opening between Graphene Dirac Cones Induced by Na Adsorption onto an Ir Superlattice.* Acs Nano, 2012. **6**(1): p. 199-204.

29. Bostwick, A., et al., *Quasiparticle dynamics in graphene.* Nature Physics, 2007. **3**(1): p. 36-40.

30. Mulazzi, M., et al., *Matrix element effects in angle-resolved valence band photoemission with polarized light from the Ni(111) surface.* Physical Review B, 2006. **74**(3).

31. Haarlammert, T., et al., *Final-state effects in photoemission experiments from graphene on Ni(111).* European Physical Journal B, 2013. **86**(5).

32. Hüfner, S., *Photoelectron spectroscopy : principles and applications.* Springer series in solid-state sciences 82. 1995, Berlin ; New York: Springer-Verlag. xii, 511 p.

33. We further confirm this statement in our supplementary material.

34. Sandin, A., et al., *Multiple coexisting intercalation structures of sodium in epitaxial graphene-SiC interfaces.* Physical Review B, 2012. **85**(12): p. 125410-124115.

35. Watcharinyanon, S., et al., *Changes in structural and electronic properties of graphene grown on 6H-SiC(0001) induced by Na deposition.* Journal of Applied Physics, 2012. **111**(8): p. 083711-083717.

36. Dedkov, Y.S., M. Fonin, and C. Laubschat, *A possible source of spin-polarized electrons: The inert graphene/Ni(111) system.* Applied Physics Letters, 2008. **92**(5).

37. Amft, M., et al., *Adsorption of Cu, Ag, and Au atoms on graphene including van der Waals interactions.* Journal of Physics-Condensed Matter, 2011. **23**(39): p. 395001-395011.

38. Ren, Y.J., et al., *Controlling the electrical transport properties of graphene by in situ metal deposition.* Applied Physics Letters, 2010. **97**(5): p. 053107-05310.

39. Chakarov, D.V., et al., *Photos induced desorption and intercalation of potassium atoms deposited on graphite(0001).* Applied Surface Science, 1996. **106**: p. 186-192.

40. McChesney, J.L., et al., *Extended van Hove Singularity and Superconducting Instability in Doped Graphene.* Physical Review Letters, 2010. **104**(13): p. 136803-136808.

41. Xia, C., et al., *Detailed studies of Na intercalation on furnace-grown graphene on 6H-SiC(0001).* Surface Science, 2013. **613**: p. 88-94.

42. Boukhvalov, D.W. and C. Virojanadara, *Penetration of alkali atoms throughout a graphene membrane: theoretical modeling.* Nanoscale, 2012. **4**(5): p. 1749-1753.

43. Hwang, C., et al., *Fermi velocity engineering in graphene by substrate modification.* Scientific Reports, 2012. **2**: p. 590-594.

44. In the experiment, all states at binding energies from -2.8 eV to -1.5 eV, which are visible in Fig. 2(a), are surpassed by the pristine π state in Fig. 2(b).

45. Gamo, Y., et al., *Atomic structure of monolayer graphite formed on Ni(111).* Surface Science, 1997. **374**(1-3): p. 61-64.

46. Sandell, A., et al., *Bonding of an isolated K atom to a surface: Experiment and theory.* Physical Review Letters, 1997. **78**(26): p. 4994-4997.

47. Watanabe, K., N. Takagi, and Y. Matsumoto, *Direct time-domain observation of ultrafast dephasing in adsorbate-substrate vibration under the influence of a hot electron bath: Cs adatoms on Pt(111).* Physical Review Letters, 2004. **92**(5): p. 057401-057405.

# Supplementary Material

**Preparation of graphene on Ni(111)**

The Ni(111) substrates were prepared by sputtering 500 nm of nickel onto α-Al$_2$O$_3$(0001) substrates as described in reference [1]. These substrates were transferred into a UHV chamber and cleaned by cycles of annealing at a temperature of 1100 K, followed by mild argon ion sputtering at a temperature of 1100 K and at a beam-energy of 1 keV. Low-energy electron diffraction (LEED) measurements exhibit a clear (111) nickel structure. The surface on Ni(111) was subsequently graphitized through catalytic dissociation of C$_2$H$_4$ at 900 K, as described in reference [2]. The deposited carbon layer was then dissolved into the Ni(111) film using a series of repeated flashes to a temperature of 1100 K, and subsequently re-segregated on the surface by cooling to room temperature at rates of ca. 8-16 K/min [3]. Note that the solubility of carbon in nickel at 1100 K is approximately 3 times higher than at 900 K. This procedure yields very high quality single layer graphene, with estimated domains of approximately 100 μm in size [4]. The quality of the nickel and graphene surfaces was verified using a combination of LEED, temperature-programmed desorption, and ARPES diagnostics.

The graphene samples were subsequently doped with Na deposited from a commercial source (SAES Na/FT/1.7/12/FT) at a pressure of ≈10$^{-10}$ Torr and at room temperature [5]. Careful annealing of the samples can then modify the adsorbate in two distinct ways. First, heating to temperatures between 300 K and 400 K for about one hour induces consolidation of the Na atoms into a uniform sub-monolayer. No significant desorption of Na or intercalation to underneath graphene was observed during consolidation, as determined from oxygen exposure (see below) and work function measurements. Second, annealing/heating the Na/Gr/Ni(111) sample above 400 K results in rapid changes in the electronic structure measured by ARPES that are due to intercalation of the Na to underneath graphene-on-Ni(111).

The ARPES spectra were measured using VUV radiation at 40.8 eV from a He discharge lamp (Specs UVS300, He IIα line, unpolarized, and a spot size of ca. 400 μm in diameter) and a hemispherical analyzer (Specs Phoibos 100) [6]. The electron pass energy was chosen to allow simultaneous detection of high and low kinetic energy electrons, from the Fermi level down to the energy of the σ$_2$ and σ$_3$ states. The energy resolution of approximately 300 meV was determined by analyzing the Fermi edge at room temperature.

**DFT calculations**

All calculations were performed using the VASP code [7] based on spin-polarized density functional theory (DFT). The projector augmented wave potentials were used for electron-ion interactions [8, 9]. Van der Waals dispersion forces were accounted for through the optB88-vdW functional by using the vdW-DF approach [10-12]. The 2p electrons of Na were explicitly treated as valence electrons by using the semi-core potential. The electron wave function was expanded using plane waves with an energy cutoff of 400 eV. The Ni(111) surface was modeled by a six layer slab separated from its periodic image in z-direction by a vacuum of at least 13 Å. The graphene was adsorbed on the Ni(111) surface by applying the commensurate condition. A (1×1) primitive cell in the top-fcc configuration (one carbon atop of Ni and the other carbon above the fcc hollow site) was used to model the graphene/Ni(111) system. To model the Na adsorption on graphene/Ni(111) or intercalation, a single Na atom was adsorbed on top of graphene or was intercalated between graphene and Ni(111) in (√3×√3) and (2×2) surface unit cells, corresponding to a Na coverage of 1 monolayer and 0.75 monolayer, respectively. The Brillouin zones of (1×1), (√3×√3), and (2×2) were sampled using (17×17×1), (11×11×1), and (9×9×1) Monkhorst-Pack k-point mesh,[13] respectively. The three bottom-most Ni(111) layers were fixed during relaxation. All structures were fully relaxed until the Hellmann-Feynman forces acting on the atoms were smaller than 0.01 eV/Å. The dipole correction was included and the electrostatic potential was adjusted accordingly [14]. To investigate the effect of Na adsorption and intercalation on the graphene states, the band structures of the Na/Gr/Ni(111) and Gr/Na/Ni(111) systems were calculated in a larger (2×2) supercell. This made it possible to compare the experimental ARPES data with the calculated band structures without unfolding the band structures of the supercell.

**Hybridization of the π state near the K-point**

In Fig. 1(s) we show the electronic structure near the K-point using settings and exposure time that enabled seeing weak spectral features near the Fermi level. These features are due to hybridization between graphene and Ni(111) states, as discusses in the text. Figure shows that the intensity of the π state is strongly diminished near the K-point as a result of hybridization. In addition, due to hybridization the state is split, as discussed in the text. The hybridized π state is visible at the K-point and near the Fermi level in figure and is consistent with results of DFT calculations.

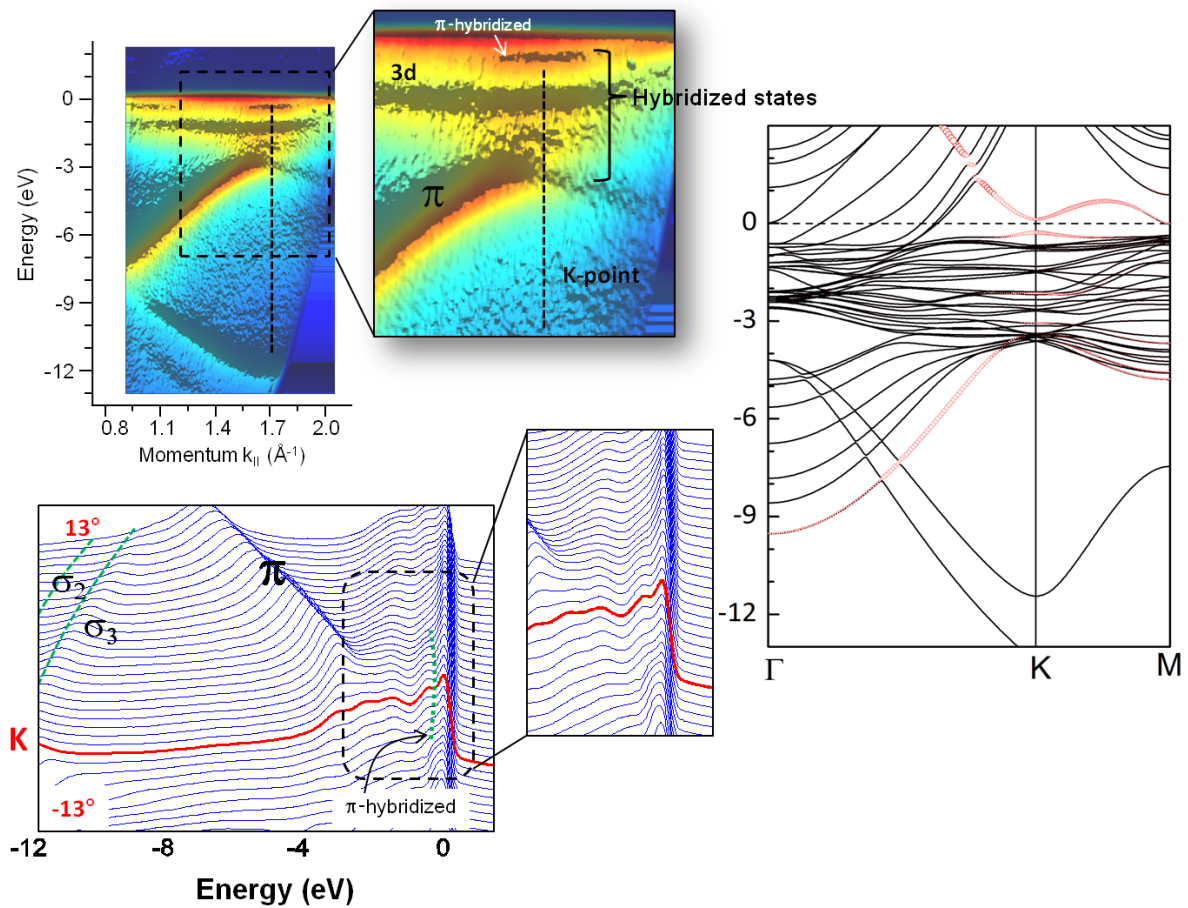

**Fig. S1.** Hybridization of the π state near the K-point in Gr/Ni(111). Upper panels show the same ARPES spectrum as in Fig.1 plotted using a different setting to show the low intensity features near the K-point. Black vertical line

(dashed) indicates position of the K-point, as established from the minimum of the σ₃ state of graphene; a manifold arising from hybridization of the 3d and π states is indicated with a curly bracket. Lower panels show the corresponding EDC lineouts. Graphene states are indicated with symbols and the EDC lineout at the K-point is indicated as red. Weak feature near the Fermi level is a hybridized state shown in red in Fig. 1(b) near the Fermi level. Right panel, DFT calculated states near the K-point, same as in Fig. 1(b).

**Intercalation of Na into Gr/Ni(111)**

Figure S2 shows snapshots of the π state evolution during the intercalation process. Panels (a), (b) and (c) are the same as in Fig. 2. The additional panels (b1) and (b2) show the intermediate stages of the intercalation process that are not shown in Fig. 2.

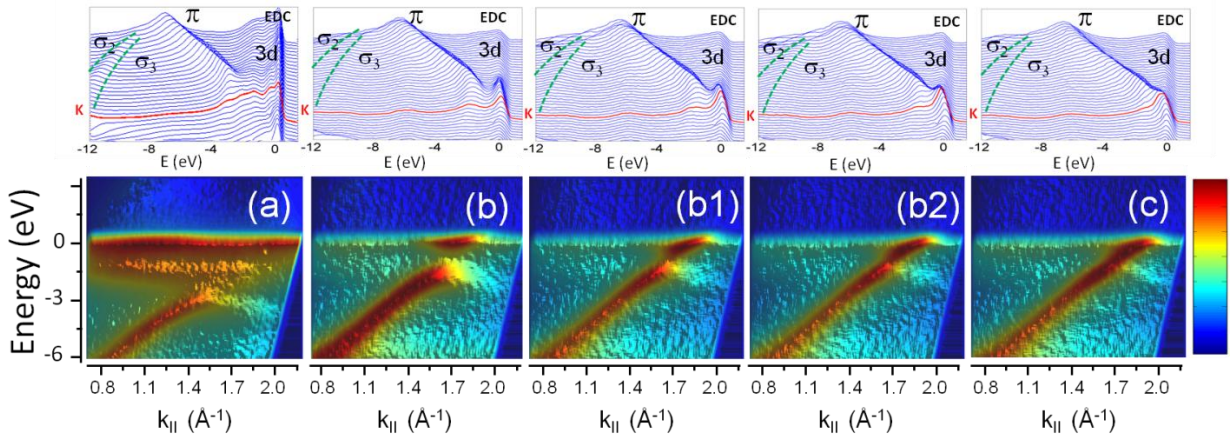

**Fig. S2.** (a) Electronic structure of Gr/Ni(111) near the K-point. (b) The same sample as in (a) but after adsorption and consolidation of 0.8 ML of Na on top. The minimum of the π* state is visible in the vicinity of the Fermi level. (b1) The same sample as in (b) during annealing; note that the π state changes. (b2) The same as in (b1) shown at a later stage of Na intercalation, with subsequent changes to the π state. (c) The same spectrum as in Fig. 2(c), the Na adatoms are intercalated into Gr/Ni(111) and decoupled the graphene layer from the substrate.

In order to confirm that the Na atoms have been intercalated, we exposed Na/Gr/Ni(111) (before intercalation) and Gr/Na/Gr/Ni(111) (after intercalation) samples to oxygen at room temperature. Na is oxidizing readily in the presence of oxygen even at the lowest concentrations [15], therefore it should oxidize if not protected by the layer of graphene. It was shown that graphene can protect intercalated atoms from oxidation, since oxygen cannot penetrate or intercalate graphene at room temperature [16, 17]. The samples remained exposed to a stream of oxygen leaked into the chamber for ≈40 minutes and we recorded ARPES spectra during this time using He II radiation. Figure S2(a) shows an ARPES spectrum recorded after adsorbing (and not intercalating) 0.8 monolayer of Na on top of Gr/Ni(111). The π* state near the Fermi level and the π-to-π* energy gap 1.3 eV indicates that the adsorbate remains on top. Two strong spectral features, near -6.5 eV and -10.5 eV, emerge after exposing sample to oxygen in UHV, shown in Fig. S3(b). These features are due to sodium oxides [15] and they indicate that the adsorbate must reside on top of graphene, being available to oxygen. After intercalation [as shown in Fig. S2(c)], we observe that no sodium oxides are formed on top of the sample after exposing the sample to oxygen for 40 min, indicating that the adsorbate is protected by graphene and does not oxidize. A very weak line at ca. -6.5 eV is visible in Fig. S3(d) that indicates trace amounts of sodium oxides formed from the very small fraction of the adsorbate that was not intercalated. The oxidation study provides undisputable proof that Na atoms are intercalated underneath Gr/Ni(111).

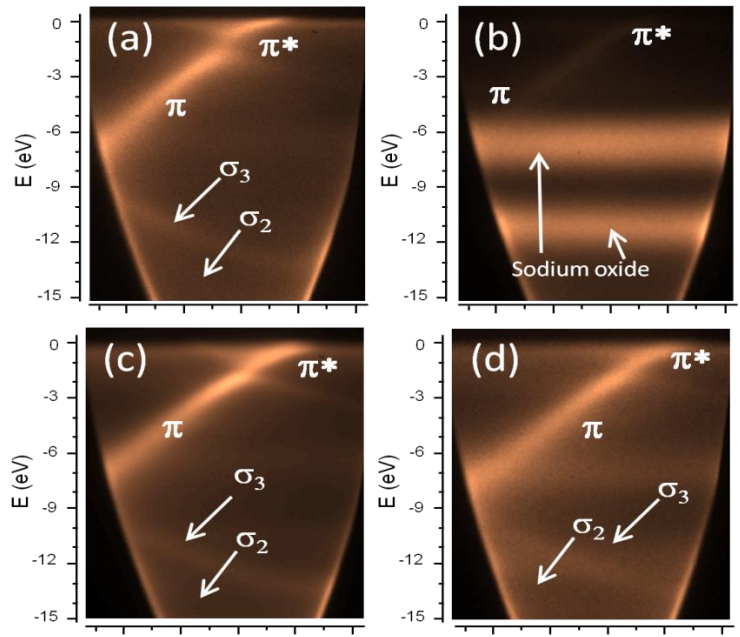

**Fig. S3.** (a) Electronic structure of Gr/Ni(111) near the K-point after adsorption and consolidation 0.8 monolayer of adsorbate on top. (b) The same sample as in (a) but after exposure to oxygen at $10^{-8}$ Torr for ca. 40 min (visible sodium oxide photoemission lines) (c) Gr/Ni(111) after adsorption and intercalation of 0.8 monolayer of Na (band gap is closed). (d) The same sample as in (c) after exposure to oxygen at $10^{-8}$ Torr for 40 min (no significant sodium oxide lines are visible).

**References**

1. Miller, D.L., et al., *Epitaxial (111) films of Cu, Ni, and CuxNiy on alpha-Al2O3 (0001) for graphene growth by chemical vapor deposition.* Journal of Applied Physics, 2012. **112**(6): p. 064317-064319.
2. Nagashima, A., N. Tejima, and C. Oshima, *Electronic States of the Pristine and Alkali-Metal-Intercalated Monolayer Graphite/Ni(111) Systems.* Physical Review B, 1994. **50**(23): p. 17487-17495.
3. Dunn, W.W., R.B. McLellan, and W.A. Oates, *SOLUBILITY OF CARBON IN COBALT AND NICKEL.* Transactions of the Metallurgical Society of America, 1968. **242**(10): p. 2129.
4. Sutter, P.W., J.I. Flege, and E.A. Sutter, *Epitaxial graphene on ruthenium.* Nature Materials, 2008. **7**(5): p. 406-411.
5. Equipment is identified in this paper only in order to adequately specify the experimental procedure. Such identification does not imply endorsement by the National Institute of Standards and Technology, nor does it imply that the equipment identified is the best available for the purpose.
6. Equipment is identified in this paper only in order to adequately specify the experimental procedure. Such identification does not imply endorsement by the National Institute of Standards and Technology, nor does it imply that the equipment identified is the best available for the purpose.

7. Kresse, G. and J. Furthmüller, *Efficient iterative schemes for ab initio total-energy calculations using a plane-wave basis set.* Phys. Rev. B, 1996. **54**: p. 11169-11186.
8. Blochl, P.E., *PROJECTOR AUGMENTED-WAVE METHOD.* Physical Review B, 1994. **50**(24): p. 17953-17979.
9. Kresse, G. and D. Joubert, *From ultrasoft pseudopotentials to the projector augmented-wave method.* Physical Review B, 1999. **59**(3): p. 1758-1775.
10. Klimes, J., D.R. Bowler, and A. Michaelides, *Chemical accuracy for the van der Waals density functional.* Journal of Physics-Condensed Matter, 2010. **22**(2): p. 022201-022206.
11. Klimes, J., D.R. Bowler, and A. Michaelides, *Van der Waals density functionals applied to solids.* Physical Review B, 2011. **83**: p. 195131-195144.
12. Dion, M., et al., *Van der Waals density functional for general geometries.* Physical Review Letters, 2004. **92**(24): p. 246401-246405.
13. Monkhorst, H.J. and J.D. Pack, *SPECIAL POINTS FOR BRILLOUIN-ZONE INTEGRATIONS.* Physical Review B, 1976. **13**(12): p. 5188-5192.
14. Neugebauer, J. and M. Scheffler, *ADSORBATE-SUBSTRATE AND ADSORBATE-ADSORBATE INTERACTIONS OF NA AND K ADLAYERS ON AL(111).* Physical Review B, 1992. **46**(24): p. 16067-16080.
15. Shek, M.L., et al., *Interaction of Oxygen with Sodium at 80-K and 20-K.* Physical Review B, 1986. **34**(6): p. 3741-3749.
16. Dedkov, Y.S., et al., *Graphene-protected iron layer on Ni(111).* Applied Physics Letters, 2008. **93**(2): p. 022509-022513.
17. Dedkov, Y.S., M. Fonin, and C. Laubschat, *A possible source of spin-polarized electrons: The inert graphene/Ni(111) system.* Applied Physics Letters, 2008. **92**(5).